\documentclass[]{aastex631}

\usepackage{natbib}
\usepackage{hyperref}

\begin{document}

\title{An Investigation of Disk Thickness in M51 from H$\alpha$, Pa$\alpha$, and Mid-Infrared Power Spectra}

\author[0000-0002-1723-6330]{Bruce~G.~Elmegreen}\affiliation{Katonah, NY 10536, USA}

\author[0000-0002-5189-8004]{Daniela Calzetti}\affiliation{Department of Astronomy, University of Massachusetts Amherst, 710 North Pleasant Street, Amherst, MA 01003, USA}

\author[0000-0002-8192-8091]{Angela~Adamo}\affiliation{Department of Astronomy, The Oskar Klein Centre, Stockholm University, AlbaNova, SE-10691 Stockholm, Sweden}

\author[0000-0002-4378-8534]{Karin~Sandstrom}\affiliation{Department of Astronomy \& Astrophysics, University of California, San Diego, 9500 Gilman Drive, La Jolla, CA 92093}

\author[0000-0002-5782-9093]{Daniel Dale}\affiliation{Department of Physics and Astronomy, University of Wyoming, Laramie, WY 82071, USA}

\author[0009-0008-4009-3391]{Varun~Bajaj}\affiliation{Space Telescope Science Institute, 3700 San Martin Drive, Baltimore, MD 21218, USA}

\author[0000-0003-4850-9589]{Martha~L.~Boyer}\affiliation{Space Telescope Science Institute, 3700 San Martin Drive, Baltimore, MD 21218, USA}

\author[0000-0002-5259-4774]{Ana~Duarte-Cabral}\affiliation{School of Physics and Astronomy, Cardiff University, The Parade, Cardiff, CF24 3AA, United Kingdom}

\author[0000-0001-8241-7704]{Ryan~Chown}
\affiliation{Department of Astronomy, Ohio State University, 180 W. 18th Ave, Columbus, Ohio 43210}

\author[0000-0001-6464-3257]{Matteo~Correnti}\affiliation{Space Telescope Science Institute, 3700 San Martin Drive, Baltimore, MD 21218, USA}

\author[0000-0002-1264-2006]{Julianne J. Dalcanton}\affiliation{Astronomy Department, University of Washington, Seattle, WA 98195, USA},
\affiliation{Center for Computational Astrophysics, Flatiron Institute, 162 Fifth Avenue, New York, NY 10010, USA}

\author[0000-0002-0846-936X]{Bruce T. Draine}\affiliation{Department of Astrophysical Sciences, Princeton University, Princeton, NJ 08544, USA}

\author[0000-0003-4224-6829]{Brandt~Gaches}\affiliation{Department of Space, Earth and Environment, Chalmers University of Technology, Gothenburg SE-412 96, Sweden}

\author[0000-0001-8608-0408]{John S. Gallagher, III}
\affiliation{Department of Astronomy, University of Wisconsin-Madison, 475 N. Charter St., Madison, WI 53706, USA}

\author[0000-0002-3247-5321]{Kathryn Grasha}\affiliation{Research School of Astronomy and Astrophysics, Australian National University, Canberra, ACT 2611, Australia}
\affiliation{ARC Centre of Excellence for All Sky Astrophysics in 3 Dimensions (ASTRO 3D), Australia}

\author[0000-0003-4910-8939]{Benjamin~Gregg}\affiliation{Department of Astronomy, University of Massachusetts Amherst, 710 North Pleasant Street, Amherst, MA 01003, USA}

\author[0000-0001-9162-2371]{Leslie K. Hunt} \affiliation{INAF -- Osservatorio Astrofisico di Arcetri, Largo E. Fermi 5, 50125 Firenze, Italy}

\author[0000-0001-8348-2671]{Kelsey E. Johnson}\affiliation{Department of Astronomy, University of Virginia, Charlottesville, VA, USA}

\author[0000-0001-5448-1821]{Robert Kennicutt, Jr.}\affiliation{Department of Physics and Astronomy, Texas A\&M University, 578 University Drive, College Station, TX 77843, USA},
\affiliation{George P. and Cynthia W. Mitchell Institute for Fundamental Physics \& Astronomy, Texas A\&M University, 578 University Drive, College Station, TX 77843, USA},
\affiliation{Department of Astronomy and Steward Observatory, University of Arizona, 933 N. Cherry Avenue, Tucson, AZ 85721, USA}

\author[0000-0002-0560-3172]{Ralf S.\ Klessen}\affiliation{Universit\"{a}t Heidelberg, Zentrum f\"{u}r Astronomie, Institut f\"{u}r Theoretische Astrophysik, Albert-Ueberle-Str.\ 2, 69120 Heidelberg, Germany}\affiliation{Universit\"{a}t Heidelberg, Interdisziplin\"{a}res Zentrum f\"{u}r Wissenschaftliches Rechnen, Im Neuenheimer Feld 225, 69120 Heidelberg, Germany}\affiliation{Harvard-Smithsonian Center for Astrophysics, 60 Garden Street, Cambridge, MA 02138, USA}\affiliation{Elizabeth S. and Richard M. Cashin Fellow at the Radcliffe Institute for Advanced Studies at Harvard University, 10 Garden Street, Cambridge, MA 02138, USA}

\author[0000-0002-2545-1700]{Adam~K.~Leroy}\affiliation{Department of Astronomy, The Ohio State University, 4055 McPherson Laboratory, 140 West 18th Avenue, Columbus, OH 43210, USA}

\author[0000-0002-1000-6081]{Sean~Linden}\affiliation{Department of Astronomy and Steward Observatory, University of Arizona, 933 N. Cherry Avenue, Tucson, AZ 85721, USA}

\author[0000-0002-5456-523X]{Anna~F.~McLeod}\affiliation{Centre for Extragalactic Astronomy, Department of Physics, Durham University, South Road, Durham DH1 3LE, UK}\affiliation{Institute for Computational Cosmology, Department of Physics, Durham University, South Road, Durham DH1 3LE, UK}

\author[0000-0003-1427-2456]{Matteo~Messa}\affiliation{INAF—Osservatorio di Astrofisica e Scienza dello Spazio di Bologna, Via Gobetti 93/3, I-40129 Bologna, Italy}

\author[0000-0002-3005-1349]{G\"oran~\"Ostlin}\affiliation{Department of Astronomy, The Oskar Klein Centre, Stockholm University, AlbaNova, SE-10691 Stockholm, Sweden}

\author[0000-0002-3472-0490]{Mansi~Padave}
\affiliation{Department of Astronomy and Astrophysics, University of California, San Diego, 9500 Gilman Drive, La Jolla, CA 92093, USA}

\author[0000-0001-6326-7069]{Julia Roman-Duval} \affiliation{Space Telescope Science Institute, 3700 San Martin Drive, Baltimore, MD 21218, USA}

\author[0000-0003-1545-5078]
{J.D.~Smith}\affiliation{Ritter Astrophysical Research Center, University of Toledo, Toledo, OH 43606, USA}

\author[0000-0003-4793-7880]{Fabian~Walter}\affiliation{Max-Planck-Institut für Astronomie, Königstuhl 17, D-69117, Heidelberg, Germany}

\author[0009-0005-8923-558X]{Tony~D.~Weinbeck}
\affiliation{Department of Physics and Astronomy, University of Wyoming, Laramie, WY 82071, USA}

\begin{abstract}
Power spectra (PS) of high-resolution images of M51 (NGC 5194) taken with the Hubble Space Telescope and the James Webb Space Telescope have been examined for evidence of disk thickness in the form of a change in slope between large scales, which map two-dimensional correlated structures, and small scales, which map three-dimensional correlated structures. Such a slope change is observed here in H$\alpha$, and possibly Pa$\alpha$, using average PS of azimuthal intensity scans that avoid bright peaks.  The physical scale of the slope change occurs at $\sim120$ pc and $\sim170$ pc for these two transitions, respectively.  A radial dependence in the shape of the H$\alpha$ PS also suggests that the length scale drops from $\sim180$ pc at 5 kpc, to $\sim90$ pc at 2 kpc, to $\sim25$ pc in the central $\sim$kpc. We interpret these lengths as comparable to the thicknesses of the star-forming disk traced by HII regions.  The corresponding emission measure is $\sim100$ times larger than what is expected from the diffuse ionized gas. PS of JWST Mid-IR Instrument (MIRI) images in 8 passbands have more gradual changes in slope, making it difficult to determine a specific value of the thickness for this emission.
\end{abstract}


\section{Introduction} \label{sec:intro}

The thickness of a face-on galaxy may be inferred from the Fourier Transform Power Spectrum (PS) of turbulent gas, appearing as a transition where a shallow slope from two-dimensional structure on large scales changes to a steeper slope from three-dimensional structure on small scales \citep{lazarian00}. Numerical simulations \citep{bournaud10,combes12,fensch23} illustrate this transition by showing large-scale velocities with two components in the plane of the disk, and small scale velocities with three components inside the disk. PS breaks at the disk thickness have been observed previously for three galaxies, the Large Magellanic Cloud (LMC), M33, and NGC 1058.  The LMC shows the PS break in HI \citep{elmegreen01} and far-infrared dust emission \citep{block10}, with inferred disk thicknesses ranging from $\sim50$ pc near 30 Dor and other bright HII regions, to $\sim250$ pc in the central regions and $\sim400$ pc in the outer regions \citep{szot19}.  In M33, the scale of the PS break ranges from $\sim50$ pc in the FUV and NUV, to $\sim100$ pc in HI and CO, to $\sim300$ pc in H$\alpha$, with the same full range also in the infrared \citep{combes12}. In NGC 1058, which is the furthest of these at a distance of $\sim10$ Mpc, the transition was seen at $490\pm90$ pc in HI emission \citep{dutta09}.  

The purpose of this paper is to report a fourth thickness measurement from PS, 
including an increase in thickness with radius, using  0\farcs07-resolution H$\alpha$ and Pa$\alpha$ images of M51 \cite[2.54 pc FWHM at 7.5 Mpc,][]{csornyei23}. We also examine PS of MIRI images of M51 at 0\farcs2 to 0\farcs6 resolution (7.3 pc to 22 pc FWHM). Thickness is important for converting surface density into volume density for dynamical considerations \citep[e.g.,][]{peng88,bacchini20}, correcting an observed rotation curve for radial pressure gradients to determine radial forces and the density of dark matter \citep{verbeke17}, and using it with the vertical velocity dispersion to determine the total disk mass \citep[e.g.,][]{sarkar18}.

The thickness of M51 has been measured before. \citet{pety13} combined the Plateau de Bure interferometer and the IRAM 30-m telescope to infer that there are two molecular components, one with a scale height of 190-250 pc containing about half the $^{12}$CO flux and another with a scale height of $\sim40$ pc and an average density $\sim10$ times higher. They also suggested the HI in M51 has a similar two-tiered structure, based on a two-component fit to the average emission line profile. \citet{hu13} derived a stellar scale height of 95 pc to 178 pc in M51 using the spiral arm pitch angle and density wave theory. A more direct comparison with our work was by \citet{tress20}, who found a break in the PS of molecular gas at a scale of around 80 pc using numerical simulations of M51.

The PS method for determining thickness is not always possible.  
\citet{koch20} examined PS of mid- to far-infrared dust emission from the LMC, SMC, M31 and M33, and HI and CO emission from M31 and M33, finding no breaks other than what might result from bright point sources viewed through a point spread function (PSF) and the exponential radial profile of the disk.  We also saw a dominant influence of point sources and exponential profiles in \citet[][hereafter Paper I]{elmegreen25}, which used mid-infrared JWST data to examine PS for NGC 628, NGC 5236, NGC 4449, and NGC 5068.  We get around those problems here by using intensity scans in the azimuthal direction and selecting only scans without strong point sources. 

Observations of other galaxies, usually with HI, do not generally have high enough spatial resolution to see a break in the PS from disk thickness.  \citet{dutta10} measured a PS slope of $-1.7$ for NGC 4254, but could only observe scales larger than 1.7 kpc. In a second study, \citet{dutta13} measured slopes averaging $-1.3$ for 18 other spirals, but for scales larger than the likely thicknesses.  The HI column density in NGC 6946
has a PS slope of $-0.96\pm0.05$  determined by \citet{nandakumar23} down to $\sim150$ pc. \citet{grisdale17} measured PS of the HI surface densities for six THINGS galaxies, finding large-scale slopes of $-2.2$ for NGC 628, $-2.8$ for NGC 3521, $-2.1$ for NGC 4736, $-2.2$ for NGC 5055, $-2.5$ for NGC 5457, and $-1.6$ for NGC 6946. All of these PS are for two-dimensional maps. The PS determined here are from one-dimensional intensity scans, which are used to avoid bright point-like sources.  The PS slope for a one-dimensional scan should be shallower by 1 than the slope for a two-dimensional scan. 

For the PS method to work, the physical scale of the thickness has to be resolved by at least a factor of $\sim10$, and the thickness has to be smaller than the whole galaxy by another factor of $\sim10$, in order to determine power law slopes on either side of the PS break. This dual constraint is satisfied for HI and far-infrared observations of the angular-largest galaxies in the local group (the LMC and M33), and it is satisfied by ground-based, optical observations of large galaxies out to $\sim3$ Mpc, such as M81 \citep{elmegreen03}, but more distant spirals are expected to have disks that are too thin in angle to resolve in the first constraint, and nearby dwarfs, including the SMC \citep{szot19}, are probably too thick to satisfy the second constraint \citep[e.g.][]{patra20a}. A third constraint is that the emission has to come from spatially-correlated structures so the PS is a power law and not, for example, flat on a log-log plot from noise. Such correlated structure was an early discovery for HI gas \citep{crovisier83}, showing that it is highly turbulent \citep[see also][]{armstrong95,elmegreen04}.

The thickness of a gas disk can also be measured using theoretical considerations of vertical equilibrium \citep[e.g.,][]{spitzer42,parker66,vanderkruit81,narayan02,koyama09,ostriker10}. This requires, for the upward component of force, observations of the gas velocity dispersion combined with estimates of magnetic and cosmic ray pressures, and, for the downward force, observations of the total mass surface density from gas, stars and dark matter inside the gas disk, along with the vertical component of forces from the bulge and dark matter halo \citep[e.g.,][]{girichidis16,hill18,wilson19}. 

The equilibrium method has been applied to the HI layers of many galaxies, including the Milky Way \citep{kalberla07,banerjee11a} and outer parts of M31  \citep{banerjee08}, 20 galaxies in \citet{bagetakos11}, 4 dwarf irregulars in \citet{banerjee11b}, superthin galaxies in \citet{banerjee10} and \citet{banerjee13}, 20 dwarf irregulars in \citet{elmegreen15}, 7 spirals and 23 dwarfs in \citet{patra20b} and \citet{patra20a}, 10 dwarf irregulars in \citet{bacchini20}, 28 HI-rich galaxies and 26 comparison galaxies in \citet{rand21}, and an ultra diffuse galaxy plus 14 dwarf irregulars in \citet{li22}.  Disk thicknesses for HI typically exceed several hundred pc.
\citet{bacchini19} determined HI and molecular thicknesses in 12 spirals, while for the molecular component alone, \citet{patra19} measured 8 spirals, \citet{wilson19} 5 ULIRGs, and \citet{molina20} 2 starburst galaxies at redshift $z\sim0.15$. 
For NGC 6946, \citet{patra21} used the equilibrium method for a multi-component disk to compare model CO line profiles with observed profiles and found that two molecular components fit best: one thin, with 30\% of the emission and a HWHM of $\sim50$ pc at 4 kpc radius, and the other twice as thick and coincident with the HI disk. 

Edge-on galaxies, including the Milky Way, have had their thicknesses measured directly.  The half-width half-maximum (HWHM) of the molecular cloud layer in the Milky Way is $\sim50$ pc inside the solar circle \citep{heyer15}. This is determined mostly by the dense molecular gas detected in both $^{12}$CO and $^{13}$CO, while the diffuse CO, detected only in $^{12}$CO, is $\sim50$\% thicker \citep{roman16}. Numerical simulations of the Milky Way by \citet{jeffreson22} obtained a similar $\sim50$ pc HWHM. NGC 891 has a CO HWHM of 110 pc \citep{scoville93} to 160 pc \citep{yim11}, which is consistent with the HWHM of 105 pc for 133 8-$\mu$m star-forming cores in that galaxy \citep{elmegreen20}.  NGC 891 also has a much wider layer of diffuse ionized gas with a scale height of $\sim1$ kpc \citep{rand90,dettmar90}, composed of nearly vertical filaments \citep{howk00,rossa04}. A CO counterpart to the thick disk in NGC 891 was suggested by \citet{garcia92}  but not seen by \citet{yim11}, although \citet{yim11} measured the HWHM of HI to be 435 pc for a single component fit and 325 pc and 1 kpc for a two-component fit. \citet{patra18} derived an equilibrium HWHM of 50 pc to 80 pc at 6 kpc radius for molecules in the edge-on galaxy NGC 7331. HI in the edge-on dwarf irregular  KK250 has a HWHM of 350 pc \citep{patra14}, and HI in the spiral IC 2233 has a HWHM of 500 pc \citep{matthews08}. In NGC 4157, NGC 4565, and NGC 5907, \citet{yim14} measured disk thicknesses as a function of radius for CO and HI; at 4 kpc radius, their linear fits imply HWHM of 120 pc, 45 pc, and 50 pc for CO and 450 pc, 180 pc, and 400 pc for HI, respectively (converting from their Gaussian widths to HWHM). 

Whenever it can be observed, disk thickness increases with radius, although not exponentially as it would for a constant velocity dispersion and exponentially decreasing surface density. Vertical velocity dispersions tend to decrease with radius \citep{tamburro09}, making the thickness increase more linearly \citep[e.g.,][]{yim11,yim14,elmegreen20}.  

The organization of this paper is as follows, Section \ref{sect:data} summarizes the H$\alpha$, Pa$\alpha$, and broadband data used here, Sections \ref{sect:ha}, \ref{sect:pa}, and \ref{sect:bb} show the power spectra for H$\alpha$, Pa$\alpha$, and 8 mid-infrared bands, Section \ref{sect:discussion} contains a discussion, and Section \ref{sect:conclusions} has our conclusions.

\section{Data}
\label{sect:data}

HST wide field camera (WFC) optical imaging of M51 for H$\alpha$ was obtained by GO–10452 (P.I. S. Beckwith), as part of the HST image release program by the Hubble Heritage Team. The imaging is a $2\times3$ mosaic covering the bright area of M51 and its companion NGC 5195.  Observations in F555W, F658N and F814W filters were retrieved from the MAST Archive\footnote{MAST: Mikulski Archive for Space Telescopes at the Space Telescope Science Institute; https://archive.stsci.edu/.}, reduced, and aligned to the Gaia reference system. The final pixel scale of the drizzled HST mosaics is 0\farcs04 px$^{-1}$. Flux  calibration is in units of counts s$^{-1}$, which were converted to physical units using the PHOTFLAM image header keywords.

JWST near–IR imaging with NIRCam for Pa$\alpha$ was obtained via the Cycle 1 JWST program \#1783 (Feedback in Emerging extrAgalactic Star clusTers, JWST–FEAST, P.I.: A. Adamo). For this work, we only utilize the F150W, F187N, and F200W filter mosaics. The NIRCam mosaics were processed through the JWST pipeline version 1.12.5 (Dec 2023 release) using the Calibration Reference Data System (CRDS) context ``jwst 1174.pmap.''\footnote{\label{note1}https://jwst-pipeline.readthedocs.io/en/latest/jwst/user\_documentation/reference\_files\_crds.html} The NIRCam mosaics have been aligned to the Gaia reference system with a common pixel scale of 0\farcs04, and are in units of Jy/px.

The NIRCam F187N centered on the Pa$\alpha$ line emission ($\lambda=1.8789$~$\mu$m at the fiducial redshift $z=0.00174$\footnote{\label{note0}From NED, the NASA Extragalactic Database.}) and the ACS/WFC F658N centered on the H$\alpha$+[NII] doublet line emission ($\lambda=0.6559$~$\mu$m, 0.6574~$\mu$m, 0.6595~$\mu$m at $z=0.00174$\footref{note0}), were used to derive emission line maps. The stellar continuum image for the F187N was derived by interpolating between F150W and F200W. As the F200W contains the Pa$\alpha$ emission, we iteratively subtracted the line from this filter, using the procedure described in \citet{messa21}, \citet{gregg24}, and  \citet{calzetti24}, until differences between two subsequent iterations are $\le0.1$\% in flux. The stellar continuum image to subtract from the F658N filter was derived by interpolating between F555W and F814W. F555W includes the [OIII]($\lambda 0.5007$~$\mu$m) line emission, but \citet{calzetti24} showed that in metal–rich galaxies this contribution is small, affecting the stellar continuum by $\le1.5$\%.

The $3\sigma$ detection limits for the emission lines were derived after converting the line maps into physical flux maps. The F658N map used here, which includes the H$\alpha$ and [NII] emission lines, were multiplied by the filter width of 0.0087$\mu$m\footnote{\label{note2}https://etc.stsci.edu/etcstatic/users\_guide/appendix\_b\_acs.html; https://jwst-docs.stsci.edu/jwst-near-infrared-camera/nircam-instrumentation/nircam-filters} after the continuum was subtracted. The F187N map, which contains the Pa$\alpha$ emission line, was multiplied by its filter width of 0.024~$\mu$m\footref{note2} after continuum subtraction. Both were corrected for the filter transmission curve value at the galaxy’s redshift. The resulting $3\sigma$ detection limits are then $7.7\times10^{-17}$ erg s$^{-1}$ cm$^{-2}$ arcsec$^{-2}$ for H$\alpha$ (including [NII] emission) and $3.8\times10^{-17}$ erg s$^{-1}$ cm$^{-2}$ arcsec$^{-2}$ for Pa$\alpha$.

JWST MIRI mosaics of M51 were obtained as part of both the Cycle 1 FEAST program and the Cycle 2 M51 Treasury program \#3435 (The JWST Whirpool Galaxy Treasury, P.I.: K. Sandstrom \& D. Dale). The FEAST program obtained mosaics in the F560W and F770W filters aiming to maximize overlap with the NIRCam mosaics, while the M51 Treasury obtained mosaics in the F1000W, F1130W, F1280W, F1500W, F1800W and F2100W matching the FEAST observing strategy. Both programs used a $1\times5$ pointing mosaic to cover $\sim12$ arcmin$^2$ of M51’s disk. The observations used a four-point dither cycling dither pattern. The MIRI mosaics were processed through the JWST pipeline version 1.13.4 (Feb 2024 release) using the CRDS context ``jwst 1241.pmap''\footref{note1}; these mosaics have a pixel scale of 0\farcs11 and are in units of MJy/sr.

\begin{figure}
\plottwo{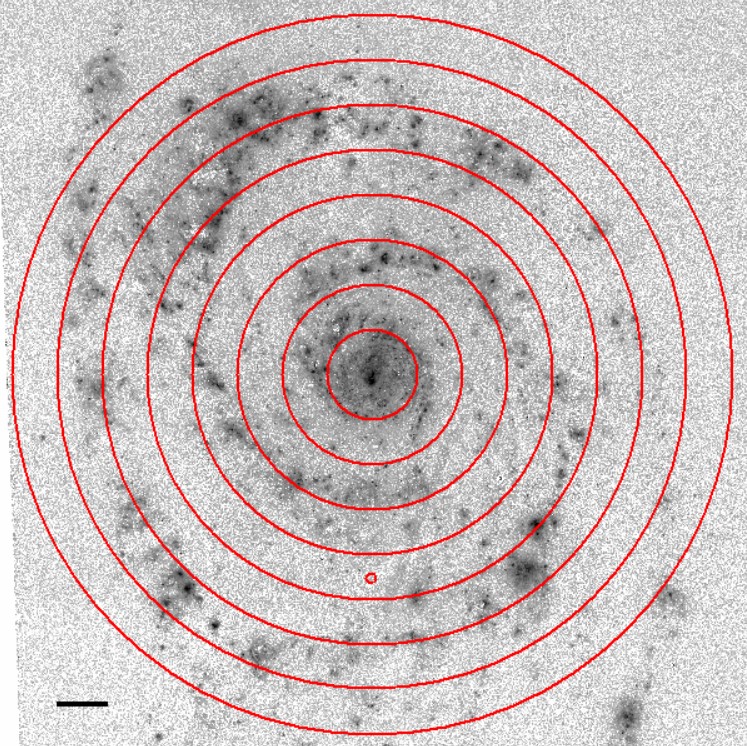}{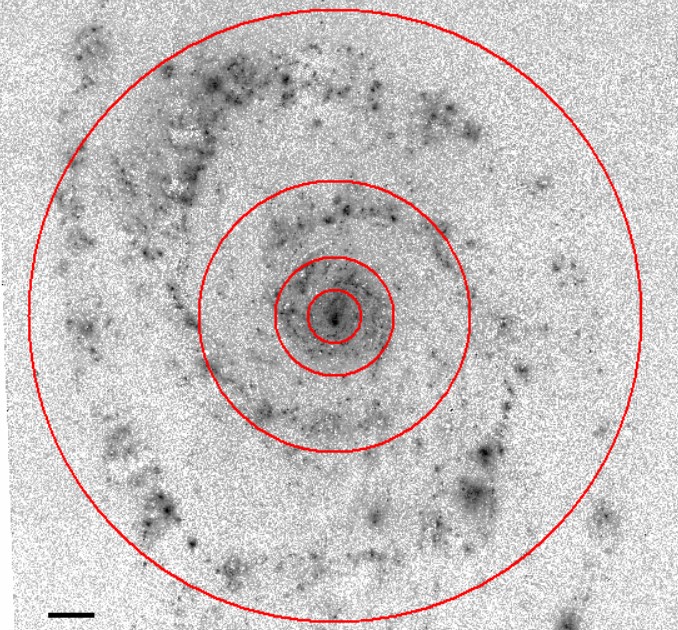}
\caption{(Left:) Image of M51 in H$\alpha$ with concentric circles at sample radii of azimuthal intensity scans used to derive the PS. The circles are spaced by 600 pixels, which is 200 scans, $24^{\prime\prime}$, or 872 pc. A small circle midway between the 4th and 5th annuli below the center has a diameter corresponding to 200 pc. (Right:) H$\alpha$ with circles at $16^{\prime\prime}$, $35^{\prime\prime}$, $80^{\prime\prime}$, and $180^{\prime\prime}$ used to determine the radial intervals for separate evaluations of thickness. The images are plotted as log of the intensity. The scale bar indicates 1 kpc. 
}
\label{fig:hacircles}
\end{figure}

\section{Power Spectra for H$\alpha$ Emission}
\label{sect:ha}

The radial profile of intensity from the exponential disk of M51 contributes to the two-dimensional PS on large scales (Paper I). To avoid that component here, PS were determined from azimuthal intensity scans.  The PS of an azimuthal scan is not exactly the same as the PS of a straight-line scan because the slight curvature of the arc inside each azimuthal pixel distorts the wavenumber scale by a small amount. The difference is unimportant if the circumference is large compared to the PS features of interest. 
Here we highlight   
$\leq200$ pc scales, which  correspond to less than 5\farcs5 or 138 pixels, which is much smaller than the circumference scan lengths ($\sim15,000$ pixels at mid-radius).  We avoid unnecessary pixel interpolation by using projected circles rather than deprojected circles in the plane of the disk. This approximation is acceptable because the $20^\circ$ inclination of M51 \citep{hu13} is small, so the line of sight thickness increases with the tilt by only $1/\cos 20^\circ=1.06$.   Also, it affects mostly the $k=2$ component of the PS by foreshortening the pixels on the major axis compared to the minor axis, and by sampling slightly different galactocentric radii on the two axes. These two effects amount to $\leq 10$\% error at $k\sim2$. This is far from the wavenumber range of interest, which is larger by the ratio of the diameter of the galaxy to the thickness, a factor of $\sim100$. 

Figure \ref{fig:hacircles} shows the positions of 8 circular scans on the H$\alpha$ image, separated uniformly by 600 pixels (872 pc). We measured the intensity in 1623 such circles, one pixel wide and separated by 3 pixels to avoid overlaps.  Figure \ref{fig:hacircles} shows every 200th scan. For comparison, a length of 200 pc corresponds to the diameter of the tiny circle between the 4th and 5th large circles below the center; it is very small on this image. The black line in the lower left corresponds to 1 kpc.   Most of the obvious features in the figure are larger than the disk thickness derived in this paper.

\begin{figure}
\begin{center}
\includegraphics[width=12cm]{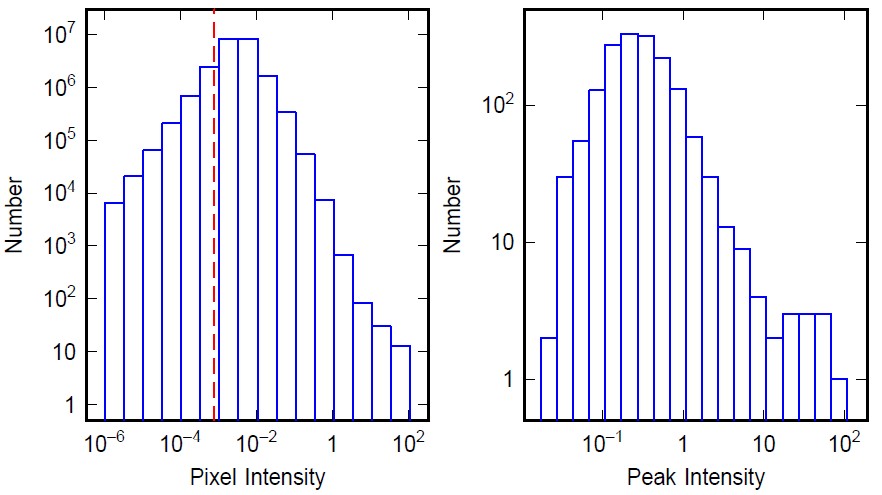}
\caption{Histograms of pixel H$\alpha$ intensity values (in units of $10^{-13}$ erg s$^{-1}$ cm$^{-2}$ arcsec$^{-2}$, left) and peak intensities in each scan (right). The peak intensity upper cutoffs used to select scans without strong sources are 0.108, 0.162, 0.216, and 0.324 in these units. These values on the abscissa of the right-hand histogram give an indication of how many scans had to be removed (i.e., those with higher peak values) to see the break in the H$\alpha$ PS. The vertical dashed line on the left-hand histogram is the $3\sigma$ noise level. 
}
\label{fig:histogram}
\end{center}
\end{figure}

Figure \ref{fig:histogram} plots the distribution function of pixel values (left) and peak values for each scan (right). As noted above, the $3\sigma$ detection for H$\alpha$ is 
$7.7\times10^{-17}$ erg s$^{-1}$ cm$^{-2}$ arcsec$^{-2}$, which is $10^{-3.1}$ on the x-axis in Figure \ref{fig:histogram} (red dashed line). This noise limit corresponds to the maximum in the pixel intensity distribution in the left-hand panel, so the histogram at smaller values is dominated by noise. The peak intensity distribution in the right-hand panel is significantly above the noise limit.

\begin{figure}
\begin{center}
\includegraphics[width=14cm]{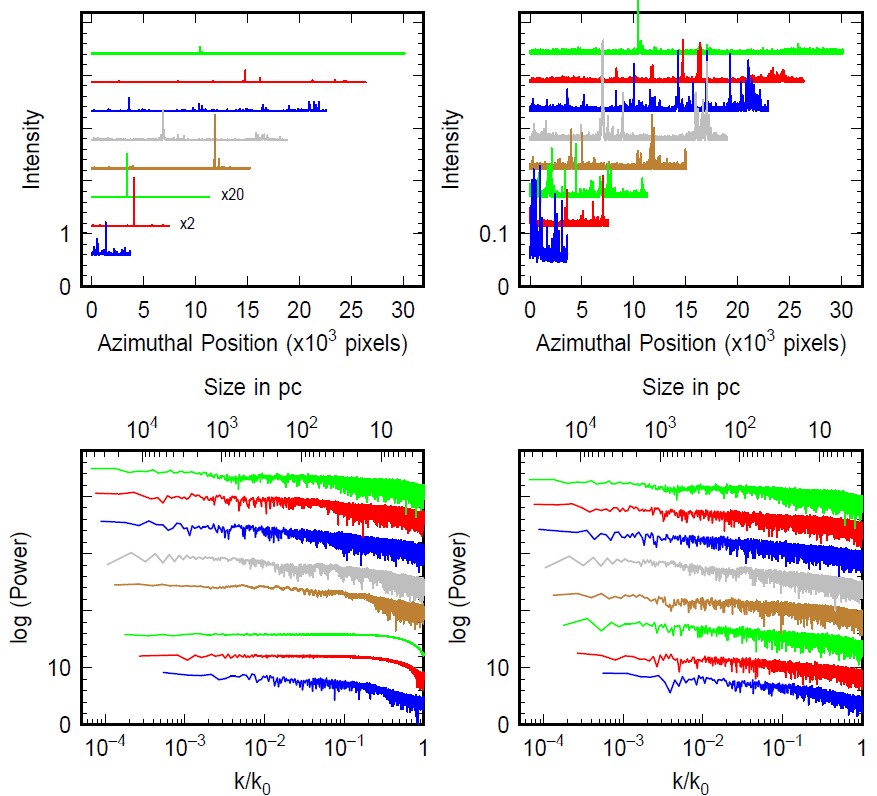}
\caption{Left: H$\alpha$ Intensity scans (with the indicated interval in units of $10^{-13}$ erg s$^{-1}$ cm$^{-2}$ arcsec$^{-2}$, top) and power spectra (in the square of these units, bottom) corresponding approximately to the positions of the circles in Fig. \ref{fig:hacircles}. Both are shifted vertically by arbitrary amounts for clarity.  The intensity ordinate is a linear scale with one unit shown; factors of 2 and 20 compressions are used for the 2nd and 3rd scans up from the bottom.  The PS ordinate shows the 10 unit scale, which means the PS covers a range of $10^{10}$ in that interval. The lengths of the scans increase with radius. Right: Intensity scans (top) and PS (bottom) also corresponding to radii near the circles in Fig. \ref{fig:hacircles}, but chosen to avoid strong sources. The PS in the lower left are irregular. For example, they are flat at low $\log k/k_0$ when there are exceptionally strong sources in the scans. When there are no strong sources, the PS (lower right) are more uniform and may be averaged to give a better composite PS. 
}
\label{fig:intensity}
\end{center}
\end{figure}

Figure \ref{fig:intensity} shows intensity scans at the top and their PS at the bottom. Those on the left are a selection of scans approximately corresponding to the 8 circles in Figure \ref{fig:hacircles}, increasing in length as the circle circumference increases. Some of these scans have single bright sources that distort the PS, as shown in the lower left. For example, the second and third scans up from the bottom have bright H$\alpha$ sources at pixel values of 4061 and 3397 respectively; the ordinate scales are compressed by factors of 2 and 20 to show these scans in the figure. The corresponding PS of these scans are relatively flat in the lower panel. To avoid including these distorted PS in radial averages, we concentrated on scans with no bright peaks, as shown in the right-hand panels. In the top-right are intensity scans close in radius to those on the left (as evident from the comparable scan lengths) but without bright peaks. Note that the ordinate stretch is larger by a factor of 10 in the top-right, and that $3\sigma$ for H$\alpha$ corresponds to 0.00077 on both ordinates. The intensity scans in the top-right have their PS in the bottom right; they are relatively similar and suitable for averaging. In what follows, we avoid H$\alpha$ intensity scans with peaks brighter than certain limits, which range between 0.108 and 0.324 in the units of the ordinate in the top plots, which is $10^{-13}$ erg s$^{-1}$ cm$^{-2}$ arcsec$^{-2}$. 

The PS at the bottom of Figure \ref{fig:intensity} plot wavenumber $k$ (inverse of length) on the abscissa, normalized to the wavenumber $k_0$ corresponding to an inverse length of 2 pixels (i.e., the number of 2-pixel wavelengths that fit in the intensity strip is $k_0$). This normalization is convenient because it does not depend on angular resolution through the PSF, which shows up as the width of the dip at high $k/k_0$, or source distance, which comes in as a coordinate conversion from $k/k_0$ to parsecs, shown on the top axis.  The PS themselves are calculated from the sum of the squares of the sine and cosine Fourier transforms using angle $2\pi kx/L$ for position $x$ from 1 to the scan length $L$ in pixels, and $k$ from 1 to $L/2$. The number of $k/k_0$ values is half the number of pixels in the scan length, and the wavelength is 2 pixels divided by $k/k_0$. The PS extend to lower $k/k_0$ values as the scan length increases. The absolute normalization of the PS on the ordinate of Figure \ref{fig:intensity} is arbitrary: they are shifted vertically for clarity. The relative scale indicated by the tick marks is accurate, so the PS slopes may be read directly from the figure using these axes. 

\begin{figure}
\begin{center}
\includegraphics[width=16cm]
{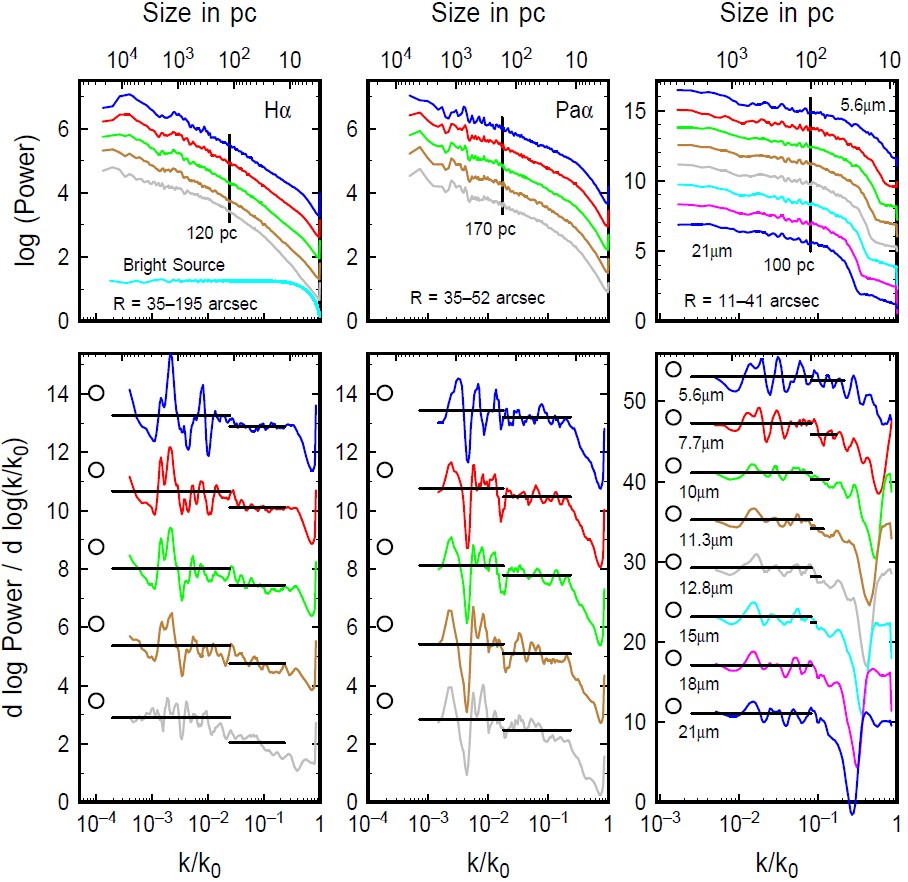}\caption{PS (top) and their running slopes (bottom) for H$\alpha$ (left), Pa$\alpha$ (middle), and eight mid-infrared passbands (right). The curves are shifted upward for clarity with slope zero-levels in the bottom panels indicated by circles.  For H$\alpha$ and Pa$\alpha$, the top curves in each panel have the lowest intensity limits and the smallest numbers of scans in the average PS, and the bottom curves in each panel have all the scans, excluding only those with negative intensity peaks. The 8 PS on the right correspond to the 8 near-infrared bands, as labeled in the bottom panel; each is from an average of the PS with the lowest intensity peaks.  A slight break at a size of $1/k\sim120$ pc is visible in the top three H$\alpha$ PS (at the vertical black line). A weaker break at $\sim170$ pc is in some of the Pa$\alpha$ PS. The mid-infrared PS on the right have a more gradual change in slope; there is no obvious break but a vertical line shows 100 pc. The running slopes in the bottom panels show sudden changes at the positions of the PS breaks. The horizontal black lines are least-square fits to the PS slopes at wavenumbers below the suggested breaks and at wavenumbers above the breaks and up to the value of $k/k_0$ corresponding to 5 times the FWHM of the PSF. The mid-infrared passbands on the right have little or no span of PS from the fiducial scale of 100 pc to the PSF. The cyan curve in the top left panel is the PS of an H$\alpha$ intensity scans with a bright point-like source.  Radial ranges for the PS are indicated in the top panel.  
}
\label{fig:derivatives}
\end{center}
\end{figure}

Figure \ref{fig:derivatives} shows PS (top) and their running derivatives (bottom) for H$\alpha$ (left), Pa$\alpha$ (middle; discussed in Sect. \ref{sect:pa}) and mid-infrared passbands (right, discussed in Sect. \ref{sect:bb}). 
The PS are averaged over the radial ranges indicated.  
The averages were made by interpolating all the individual PS, $P_i(k_i/k_{0,i})$, into values where their relative wavenumbers, $k_i/k_{0,i}$, equal the relative wavenumbers of the longest PS, $P_{\rm long}(k_{\rm long}/k_{\rm{0,long}})$, i.e., where $k_i=k_{\rm long}(k_{0,i}/k_{\rm{0,long}})$. This interpolation ensures averaging at the same physical scales, tied to pixel size. The longest PS is the one from the scan with the largest radius in the average. 

The running slopes of the average PS in the bottom panels of Figure \ref{fig:derivatives} were determined by dividing the range of $k/k_0$ into 50 equal intervals of $\log k/k_0$ and evaluating the average of the logarithm of the average PS in each interval and the average of the $\log(k/k_0)$ values in each interval. The running slope is the ratio of the difference between the interval-averaged log-PS at locally higher and lower $k/k_0$ to the difference between the interval-averaged $\log(k/k_0)$ at these higher and lower $k/k_0$. This procedure of dividing the wavenumbers into equal log intervals was done because the density of PS values on the $\log k/k_0$ axis is not constant. 
The black horizontal lines superposed on the running slopes are least-squares fits over intervals of wavenumber that are discussed below. The zero-levels for the slopes are indicated by circles. 

We consider the results for H$\alpha$ first. In order to select scans without strong sources and test the robustness of the results against the specific choice of an intensity cutoff, we experimented with 4 values that gave various numbers of PS in the average. From top to bottom in the left-hand panels of Figure \ref{fig:derivatives}, the first four PS and their corresponding derivatives exclude scans with peaks exceeding 1.08, 1.62, 2.16, and 3.24 in units of $10^{-14}$ erg s$^{-1}$ cm$^{-2}$ arcsec$^{-2}$. They also exclude scans with negative peaks, which arise from the continuum subtraction (negative intensity peaks make the same PS distortions as positive peaks). Because of these cutoffs, the top PS includes the fewest scans in the average (197 scans out of 1623 total) and has the smallest distortions from bright point sources. The fourth PS from the top includes 710 scans with slightly brighter intensity peaks. The bottom PS includes all the scans without negative peaks, even if they have large positive peaks (1130 scans). The PS most representative of the faintest emission are the top few. Large fluctuations at low $k/k_0$ are partly from noise and a low density of values there on the logarithmic axis. 

Also in the top left panel of Figure \ref{fig:derivatives}, in cyan, is an example of a PS with a bright pixel at $13^h29^m$45\farcs97, $47^\circ11^{\prime}$20\farcs56. This pixel may be an image flaw, but the scan is treated like the others so it shows the PS of the PSF along a one-dimensional scan. The PS is flat at low $k/k_0$ and dips down at high $k/k_0$ where the PSF removes finer scale structure. 

\begin{figure}
\begin{center}
\includegraphics[width=16cm]
{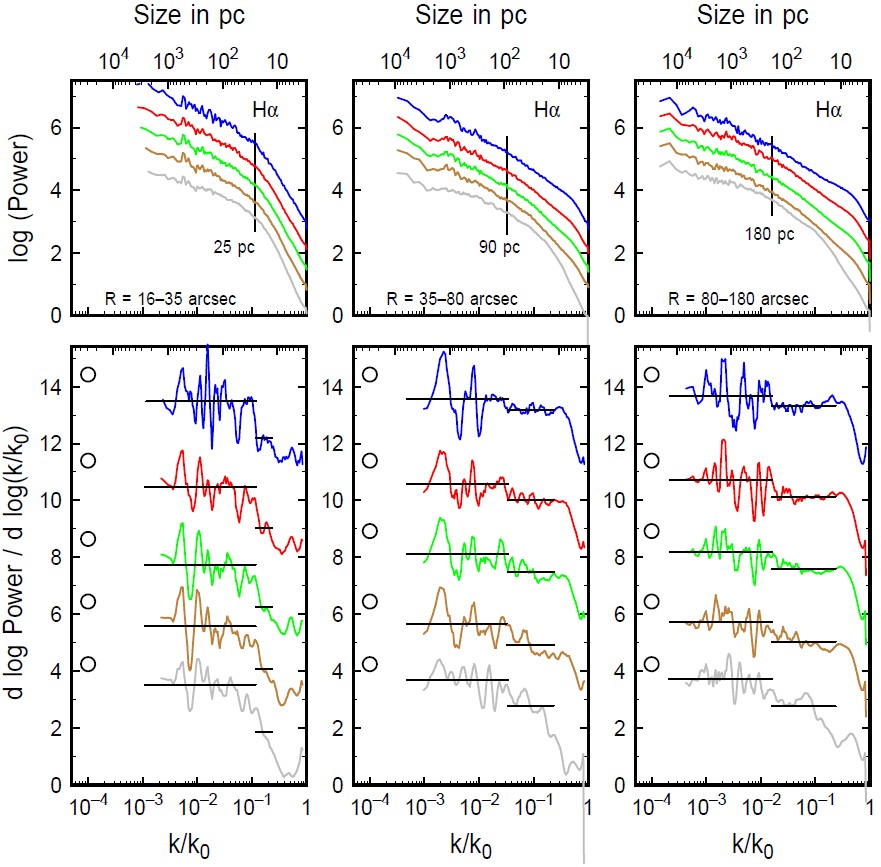}\caption{Average PS (top) and their running slopes (bottom) are shown for three indicated intervals of galactocentric radius (left to right). As for the H$\alpha$ in Fig \ref{fig:derivatives}, the peak intensities and the number of PS in the averages both increase from the top curve to the bottom curve, showing the tendency for the PS break to appear only when the weakest intensities are considered. There is a clear change in PS shape as the length scale corresponding to the PS break increases with radius. Zero-levels for the running slopes are indicated by circles. }
\label{fig:radii}
\end{center}
\end{figure}

\begin{figure}
\begin{center}
\includegraphics[width=12cm]{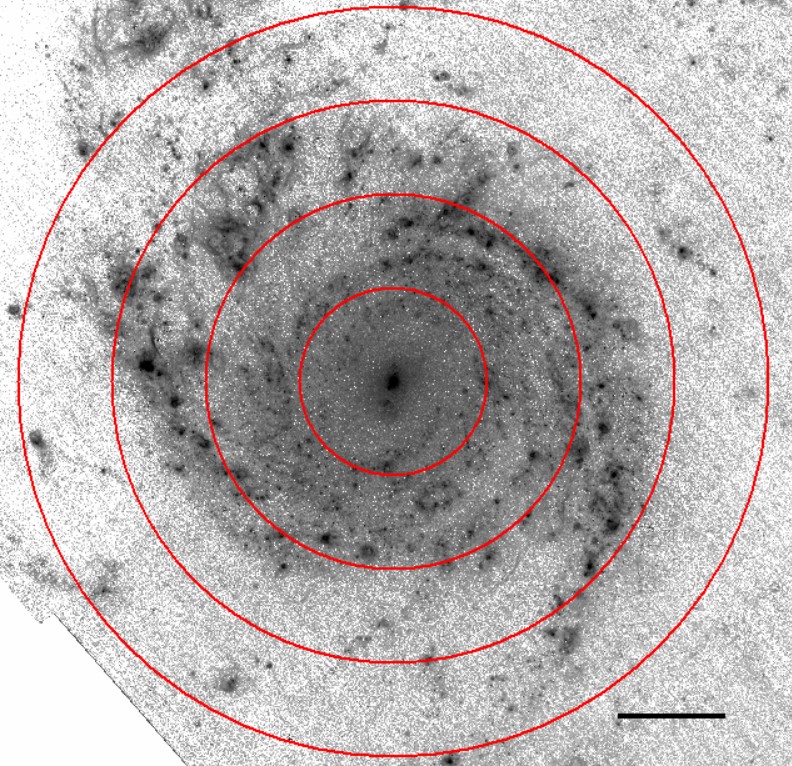}
\caption{The Pa$\alpha$ image of M51 is shown with circles corresponding to scan positions separated by 300 pixels, which is 100 scans, $12^{\prime\prime}$, and 436 pc. The image is plotted as log of the intensity. This part of the galaxy surveyed at Pa$\alpha$ is about 1/4 the size of the H$\alpha$ image in Fig. \ref{fig:hacircles}.  The scale bar is 500 pc. 
}
\label{fig:paacircles}
\end{center}
\end{figure}

The top three PS in the top left panel of Figure \ref{fig:derivatives} have slight breaks at a size scale of $1/k\sim120$ pc, which is shown by the vertical black line (as discussed in the next paragraph). The running slopes in the bottom panel show a corresponding jump at $\sim120$ pc, with approximately constant values on either side. The black horizontal lines at low $k/k_0$ are least-squares fits to the PS slopes between $k=1$ and the value of $k/k_0$ corresponding to 120 pc. The black line at large $k/k_0$ is the least-squares fit to the slope between 120 pc and the value of $k/k_0$ at 5 times the FWHM of the PSF; this FWHM for H$\alpha$ is $0.07^{\prime\prime}$, or 2.55 pc.  The intensity limits and numbers of scans included in each PS average, and the slopes for low and high $k/k_0$ parts of the PS averages, are in Table \ref{table:slopes}.  Slope errors in the table are the 90\% uncertainty limits in a student-t distribution.  The lowest PS among the 5 averages shown, which was made from the most intensity scans, has a continuously varying slope because it is a superposition of PS with different shapes (Fig. \ref{fig:intensity}).

We measured the break scale of $\sim120$ pc from the $k/k_0$ value at the intersection between the power law fits above and below a first break-scale estimate ($\sim100$ pc). These power law fits typically included a wide $k/k_0$ range and did not vary significantly if the range changed slightly. To determine the uncertainty of the break scale, we measured the rms deviation between the power law fit and the PS on each side of the break estimate, and then found the values of $k/k_0$ above and below the fitted break where the power law equaled the break value minus and plus, respectively, the rms deviations.  The physical scales corresponding to these $k/k_0$ and their uncertainties are in Table \ref{table:slopes}; 120 pc is an approximate average for these.  

The H$\alpha$ PS also suggest that the disk thickness increases with galactocentric radius. Figure \ref{fig:radii} shows three columns of average PS and their running slopes in radial intervals from $16^{\prime\prime}$ to $35^{\prime\prime}$ (0.58 kpc to 1.27 kpc) on the left, $35^{\prime\prime}$ to $80^{\prime\prime}$ (1.27 kpc to 2.91 kpc) in the middle, and $80^{\prime\prime}$ to $180^{\prime\prime}$ (2.91 kpc to 6.54 kpc) on the right. These interval limits are shown as circles on the right-hand side of Figure \ref{fig:hacircles}.  As in Figure \ref{fig:derivatives}, the numbers of scans in the average PS increase from top to bottom with a sequence of increasing cutoff values (Table \ref{table:radii}). For the mid- and large radial intervals, the cutoff sequence is the same as for H$\alpha$ in Figure \ref{fig:derivatives}, but for the lowest radial range, the cutoff values are twice as large to get enough scans in the average.

The breaks in the PS change with radius, increasing from $\sim25$ pc at $R<35^{\prime\prime}$, to $\sim90$ pc at mid-radius, to $\sim180$ pc at the larger radii (Table \ref{table:radii}). The running slopes in the bottom panels show corresponding shifts. The scale of the PS break at the smallest radius is comparable to 5 times the FWHM of the PSF, which is the limit of our capability, so the thickness there could be smaller than 25 pc. 

\newpage
\section{Power Spectrum for Paschen $\alpha$ Emission}
\label{sect:pa}

Azimuthal scans were also taken for Pa$\alpha$, using data from JWST (Section 2). The Pa$\alpha$ image with representative circles for the scans is shown in Figure \ref{fig:paacircles}.  As for H$\alpha$, the pixel size is $0.04^{\prime\prime}$ and the  scans are taken every 3 pixels. The figure shows four circles at 300-pixel, or 100-scan, intervals. All of the intensity scans were complete circles that fit inside the image, so the spatial coverage in Pa$\alpha$ is less than in H$\alpha$.

PS and PS slopes are shown in the middle of Figure \ref{fig:derivatives}. As for H$\alpha$, the curves from top to bottom are for scans with increasing limits to the peak intensity (see Table \ref{table:slopespa}). A slight break in the PS at $\sim170$ pc is evident (vertical black line), but it is more subtle than for H$\alpha$. The horizontal lines in the bottom panel show PS slopes for $k/k_0$ smaller and larger than the value at 170 pc, again extending to 5 times the FWHM. 

Compared to H$\alpha$, Pa$\alpha$ has lower signal-to-noise for the same physical regions.  For Case B recombination, the Pa$\alpha$ intensity should be $1/8$ of the H$\alpha$ intensity. The lowest threshold for Pa$\alpha$ emission in Table \ref{table:slopespa}, for which there are 43 scans, is $0.38\times10^{-14}$ erg s$^{-1}$ cm$^{-2}$ arcsec$^{-2}$. The equivalent emission measure in H$\alpha$ has an intensity 8 times larger, or $3.04\times10^{-14}$ in the same units, which is comparable to the highest H$\alpha$ threshold in Table \ref{table:slopes}, with 710 scans. 
 
\section{Power Spectra for Broadband Mid-Infrared Emission}
\label{sect:bb}

\begin{figure}
\begin{center}
\includegraphics[width=12cm]{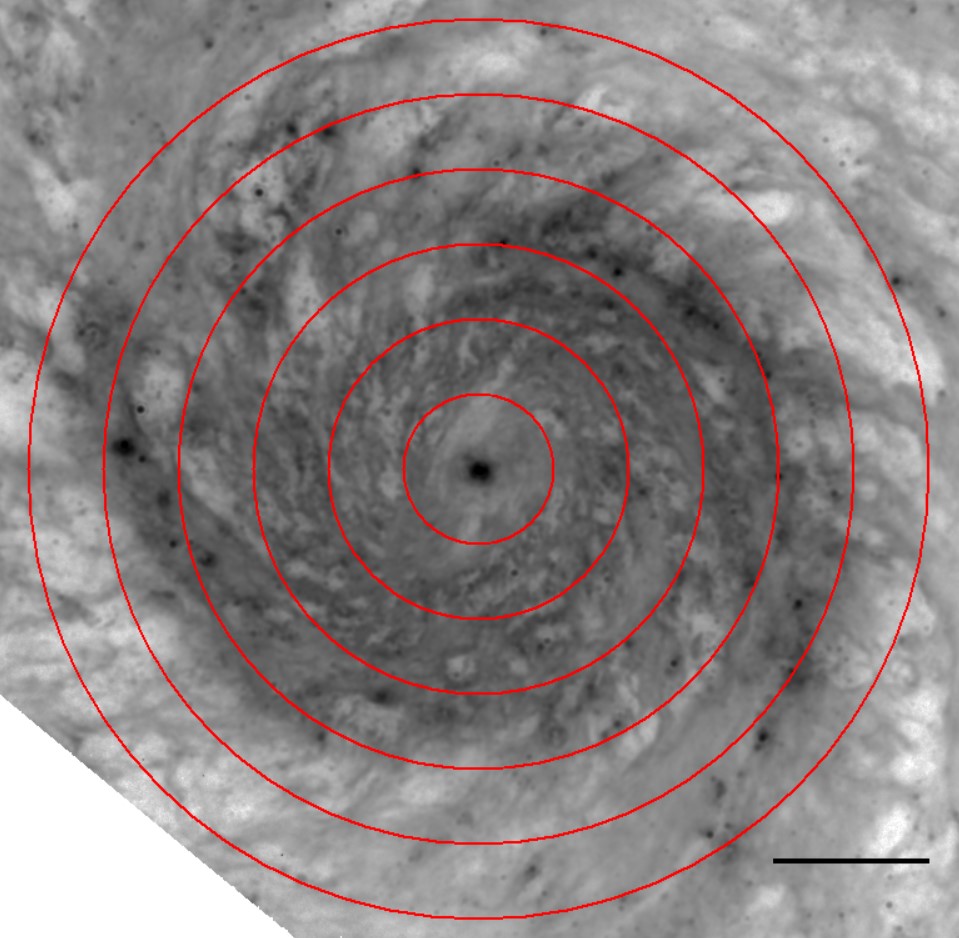}
\caption{Image of M51 in the F1130W filter with circles at the positions of sample  intensity scans spaced by 60 pixels, which corresponds to 20 scans, $6.6^{\prime\prime}$, or 240 pc. The image is plotted as log of the intensity. The scale bar is 500 pc. 
}
\label{fig:1130circles}
\end{center}
\end{figure}

Images of M51 from JWST in 8 mid-infrared passbands were examined for evidence of disk thickness in PS. Unlike the H$\alpha$ and Pa$\alpha$ images, which have $0.04^{\prime\prime}$ pixels and FWHM PSF resolutions of $0.07^{\prime\prime}$, the broadband images have $0.11^{\prime\prime}$ pixels and various resolutions, depending on the wavelength.  The (filter, FWHM resolution) combinations for the MIRI images are: (F560W, $0.207^{\prime\prime}$), (F770W, $0.269^{\prime\prime}$), (F1000W, $0.328^{\prime\prime}$), (F1130W, $0.375^{\prime\prime}$), (F1280W, $0.420^{\prime\prime}$), (F1500W, $0.488^{\prime\prime}$, (F1800W, $0.591^{\prime\prime}$), and (F2100W, $0.674^{\prime\prime}$), from the JWST User Documentation\footnote{\url{jwst-docs.stsci.edu/jwst-mid-infrared-instrument/miri-performance/miri-point-spread-functions}}.
Figure \ref{fig:1130circles} shows 6 circles representing the locations of azimuthal intensity scans separated by 20 scans (equal to 60 pixels) and superposed on the F1130W image. 

The right-hand side of Figure \ref{fig:derivatives} shows the PS and their derivatives. Each curve is for a different wavelength through the indicated sequence of JWST filters from top to bottom: F560W, F770W, F1000W, F1130W, F1280W, F1500W, F1800W, and F2100W. The increasing breadth of the dip at high $\log k/k_0$ shows the effect of larger PSFs. In all cases, only scans with low peak intensities were used, as summarized in Table \ref{table:slopesBB}. 

The broadband PS have a more gradual slope change than H$\alpha$ or Pa$\alpha$. Nevertheless, we measure the average slopes below and above the wavenumber at 100 pc (the vertical line) for comparison.  The threshold intensities, number of scans in each PS, and slopes at low and high $\log k/k_0$, are in Table \ref{table:slopesBB}.  The longest two mid-infrared wavelengths have no $k/k_0$ values below 5 times the FWHM of the PSF. 

\section{Discussion}
\label{sect:discussion}
\subsection{Power Spectrum Breaks and Slopes}

Average PS for H$\alpha$ between 1.27 kpc and 7.09 kpc radius, and Pa$\alpha$ between 1.27 kpc and 1.89 kpc radius, show a change in slope at a scale of $\sim120$ pc and $\sim170$ pc, respectively. The scale for H$\alpha$ increases with radius, from $\sim25$ pc at 0.5-1 kpc radius, to $\sim90$ pc at 1.27-2.91 kpc, to $\sim180$ pc at 2.91-6.54 kpc.  These slope changes suggest a transition from two-dimensional to three-dimensional turbulence on these scales, in which case they represent a measure of disk thickness. The radial increase in thickness is consistent with measurements in other galaxies \citep[e.g.,][]{yim14}. 

The conversion between length $1/k$ and disk thickness has not been calibrated from observations, but a discussion in \cite{bournaud10} gives some insight. They simulate an LMC-size galaxy in a 26 kpc box with a midplane resolution less than 13 pc. The average gas disk inside a radius of 5 kpc was fitted to the isothermal equilibrium function, ${\rm sech}^2(z/h)$, to obtain $h=207$ pc. We note that a Gaussian approximation to ${\rm sech}^2(z/h)$ for small $z/h$ has a dispersion of $h/\sqrt(2)$, and an exponential approximation to this function at large $z/h$ has a scale height of $2h$. The HWHM where ${\rm sech}^2(z/h)=0.5$ is at $z=0.88 h$. Also in this simulation, the PS of the surface density viewed perpendicular to the disk had a break at $k_{\rm br}\sim100$, which corresponds to $(13000 {\rm pc})/(2\pi k_{\rm br})=21.3$ pc. The authors consider a measure of thickness to be $2\pi/k_{\rm br}$, which would be 134 pc. 

\cite{combes12} measured PS breaks in M33 and in corresponding simulations. For the simulations, they compared the scale of the PS break ($1/k_{\rm br}$) and the thickness measured as $H_{\rm z} =2.35(\langle z^2 \rangle - \langle z \rangle^2)^{1/2}$, which is the FWHM for a Gaussian. They found that $k_{\rm br}H_{\rm z}$ varies between 0.24 and 1.07 with an average of 0.66 among 9 simulations, depending on the details of the feedback that contributes to turbulence on various scales. 

\cite{lazarian00} determine PS for turbulent emissions when there are bulk velocity gradients along the line of sight, as in Milky Way observations along the galactic plane. They define a characteristic scale as the length ($\lambda$ in their notation) where the line of sight turbulent speed equals the change in bulk velocity. PS show a break at $\lambda k_{\rm br} \sim1$ for observed transverse wavenumber $k_{\rm br}$. This interpretation of PS breaks is more related to the scale dependence of turbulent speed than our interpretation, which as a measure of line-of-sight thickness in another galaxy.  

Our results may be compared with measurements by \citet{pety13} of the thick component of CO in M51, which was inferred to have a scale height of 190 pc to 250 pc using combined information from radio interferometry and a single dish telescope. Our results are also comparable with the 95-178 pc thickness derived by \citet{hu13} from the spiral arm pitch angle and density wave theory. Our thinner measurements for H$\alpha$ at intermediate radii are consistent with the 40 pc scale height (80 pc thickness) for the thin component of CO in \citet{pety13} and  also with the molecular disk thickness of 80 pc in the simulation of M51 by \citet{tress20}. \cite{pety13} derive a scale height of 10 to 20 pc in the inner kpc for their thin component, which is consistent with our $\leq20$ pc measurement inside $35^{\prime\prime}$ (1.27 kpc).  Considering this, our measurement of the length scale at the break, which in our notation is $2/(k/k_0)_{\rm br}$ multiplied by the pixel size in pc (i.e., 1.45 pc for H$\alpha$), should be directly comparable to the scale height or half-thickness of the disk.

The increase in inverse break wavenumber with galactocentric radius, $R$, measured here for M51 corresponds to approximately 40 pc per kpc in radius, starting from $\sim0$ pc thickness at the center; i.e., $k_{\rm br}({\rm pc})^{-1}\sim40R({\rm kpc})$. This was determined from the $180-25=155$ pc scale increase over a radial range from 0.75 kpc to $0.5\times(2.91+6.54)=4.7$ kpc. This rate of increase for H$\alpha$ is $\sim4$ to 6 times faster than for various gas components in the edge-on galaxy NGC 891, where the scale heights have been directly measured to be $H({\rm pc})=(120\pm1)+(8.3\pm0.1)R({\rm kpc})$ for CO, $H({\rm pc})=(151\pm5)+(11.6\pm0.5)R({\rm kpc})$ for HI \citep{yim14}, and $H({\rm pc})=(49.9\pm32.9)\pm(6.3\pm3.1)R({\rm kpc})$ for bright $8\mu$m sources associated with star formation \citep{elmegreen20}. 

The fast rate of thickness increase in M51 could be related to its recent interaction with NGC 5195. Strong interactions have been found to increase the gas velocity dispersion by a factor of $\sim2$ or more in both observations \citep{elmegreen93,irwin94,kaufman97,kaufman99,goldman00,ashley13} and simulations \citep{burkhart10,renaud14}. For a minor interaction like NGC 5195 with M51, only the outer regions may be affected \citep{bournaud09}.  Because the disk thickness scales with the square of the velocity dispersion for a given surface density, the interaction could have thickened the outer parts of M51 most, increasing the rate of thickening compared to other galaxies.

The break in the PS observed here shows up most clearly for intensity scans that contain no bright point-like sources.  This differs from HI studies of nearby galaxies (LMC, M33, cf. Sect. I), which show a break for PS averaged over large sections of the disk. We could not do the equivalent two-dimensional PS or large-area averaging here because we had to avoid bright sources (trials of this type produced no breaks). The extremes of intensity for H$\alpha$, Pa$\alpha$ and MIRI passbands are much larger than for HI, and intensity scans with bright point-like sources have flat PS.

For H$\alpha$, the PS slopes at low and high $k/k_0$ average $-0.75\pm 0.03$ and $-1.28\pm0.07$, respectively, for the first four entries in Table \ref{table:slopes}. The slopes for H$\alpha$ at 0.5-1 kpc radius average $-0.95\pm0.04$
and $-2.43\pm0.26$, at 1.27-2.91 kpc they average $-0.82\pm0.05$ and $-1.42\pm0.13$. and at 2.91-6.54 kpc, they average $-0.74\pm0.04$ and $-1.29\pm0.13$. Pa$\alpha$ slopes average $-0.65\pm 0.08$ and $-0.94\pm0.08$ at low and high $k/k_0$ for the first 4 entries in Table \ref{table:slopespa}. For the broadband filters, the average slope at low $k/k_0$ is $-0.86\pm0.10$ and at high $k/k_0$ it is $-1.77\pm2.23$.

These low-$k/k_0$ slopes are all similar, averaging $-0.80\pm0.10$, but the high $k/k_0$ slopes have considerable variation, averaging $-1.52\pm0.47$. Pa$\alpha$ differs from H$\alpha$ because the difference in slopes between low and high $k/k_0$ is  smaller for Pa$\alpha$, making the presence of a break in Pa$\alpha$ more uncertain. The slope difference at mid-infrared wavelengths from the MIRI observations is larger than for either H$\alpha$ or Pa$\alpha$, but the high-$k/k_0$ slope is possibly contaminated by the steep decline of the PS into the PSF. 

\subsection{H$\alpha$ Intensities}

The low-intensity emission observed in H$\alpha$ is brighter than the diffuse ionized gas (DIG) in the Milky Way \citep{reynolds04}. The intensity values in Figure \ref{fig:histogram} and our threshold values in Table \ref{table:slopes} correspond to an intensity of $\sim10^{-14}$ erg s$^{-1}$ cm$^{-2}$ arcsec$^{-2}$. \citet{reynolds92} observed the local DIG with an intensity $I\sin(b)\sim2.9\times10^{-7}$ erg s$^{-1}$ cm$^{-2}$ sr$^{-1}$ for galactic latitude $b$, which is, after multiplying by 2 for the whole disk and converting to our units, $1.4\times10^{-17}$ erg s$^{-1}$ cm$^{-2}$ arcsec$^{-2}$.  This is $10^{-3}$ times fainter than the weak emission in M51 and fainter than our $1\sigma$ noise level. Although the star formation rate per unit area in M51 is larger than it is locally, the additional star formation is not enough to explain the M51 emission as equivalent to Milky Way DIG. 

We may get some insight into the origin of the M51 H$\alpha$ by converting the intensity to an emission measure. 
From \citet{reynolds92}, the H$\alpha$ intensity $I$ and emission measure, $EM$, for case B recombination (where Lyman emission lines are absorbed locally) are related by
\begin{equation}
I = 2.0\times10^{-18}T_4^{-0.92}EM \; {\rm erg} \;{\rm s}^{-1} {\rm cm}^{-2} {\rm arcsec}^{-2},
\end{equation}
where $EM$ is $n_e^2L$ for electron density $n_e$ in cm$^{-3}$ and pathlength L in pc. Here we have converted steradians to arcsec$^2$. Setting $T_4=0.8$ \citep{reynolds92}, and $I=10^{-14}$ erg s$^{-1}$ cm$^{-2}$ arcsec$^{-2}$ for the faint component used in our average PS, we obtain
EM$=4.0\times10^{3}$ cm$^{-6}$pc. If we use the inferred disk thickness of $L\sim100$ pc, then $n_e=6.3$ cm$^{-3}$. This electron density is much higher than in the DIG of the Milky Way, where $n_e\sim0.1$ cm$^{-3}$. 

Slit spectra measurements by \citet{hoopes03} of the DIG in M51 obtained EM$\sim30$ cm$^{-6}$pc in the interarm regions, $\sim50$ cm$^{-6}$pc in the arms and $\sim200$ cm$^{-6}$pc in a few HII regions. These correspond to H$\alpha$ intensities of 7.5, 12.6 and 50.3 in units of $10^{-17}$ erg s$^{-1}$ cm$^{-2}$ arcsec$^{-2}$, which are comparable to or slightly greater than our $3\sigma$ limit of $7.7\times10^{-17}$ erg s$^{-1}$ cm$^{-2}$ arcsec$^{-2}$.  The M51 DIG is essentially at our noise limit.

The higher emission that dominates our PS are  probably from numerous low-mass star-forming regions, while the DIG is some underlying component that does not contribute much to the PS. The average density in the thin molecular component detected by \citet{pety13} is $\sim10$ cm$^{-3}$ at mid-disk, and higher by a factor of $\sim4$ at 1 kpc. This molecular component exceeds the density of atomic hydrogen by a factor of $\sim26$ at 1 kpc \citep{pety13}. Ionized gas at this density would explain our average H$\alpha$ intensity. The presence and even dominance of such dense gas in ionized and molecular form suggest very different interstellar media in the Milky Way and M51, making it conceivable that there is a change from discrete and separate HII regions in a Milky Way type galaxy to overlapping HII regions in M51. 

Some of the H$\alpha$ emission from M51 could come from scattering off dust. \citet{barnes15} show from MHD simulations and fractal models that most of the scattering occurs near the midplane near the brightest HII regions, but 5\% to 10\% can come from a height of 300 pc. For a face-on view, less then 10\% to 20\% of the H$\alpha$ is scattered, although the scattering fraction off the surfaces of dense, high latitude clouds could be $\sim40$\%. 

Our previous observations of four other galaxies in several of the mid-infrared JWST bands used here, including the spirals NGC 628 and NGC 5236, also showed no break in the PS, even when considering the faintest emission (Paper I). We suggested that the disks emitting in these passbands are too thin for the spatial resolution. The same could be true for the MIRI passbands here.

\section{Conclusions}
\label{sect:conclusions}
The power spectrum of H$\alpha$ emission from M51 has a sudden change of slope at a scale of $\sim120$ pc over the main part of the disk, with a range in these scales from $\sim25$ pc in the central kpc to $\sim90$ pc at $\sim2$ kpc and $\sim180$ pc at 5 kpc. These slope changes are interpreted as measures of the line-of-sight disk thickness.  A less certain measurement at Pa$\alpha$ gives a thickness of $\sim170$ pc over a radial range of $\sim1.3$ kpc to $\sim1.9$ kpc.

The characteristic intensity of H$\alpha$ emission corresponds to an emission measure of $4.0\times10^{3}$ cm$^{-6}$pc, and, for a line-of-sight thickness of 100 pc, an electron density of $6.4$ cm$^{-3}$. This emission measure is nearly two orders of magnitude higher than the DIG emission measure in the Milky Way and M51, which emphasizes that our PS are dominated by star-forming regions. 

Broadband emission in the JWST MIRI filters did not indicate a particular length scale in the PS, although the slope clearly changes from large scales to small. Part of the difference with the other observations is that the PSF is broad at these mid-infrared wavelengths. In addition, the mid-infrared emission from dust could trace a thinner component of the gas. 

\newpage

\begin{deluxetable*}{cccccc}
\tabletypesize{\scriptsize}
\tablecaption{Power Spectrum Results for H$\alpha$ at Various Peak Intensity Thresholds\label{table:slopes}}
\tablewidth{0pt}
\startdata
&&&&&\\
&&&&&\\
Threshold\tablenotemark{a}&0.108&0.162&0.216&0.324&none\\
Number of scans& 197&   375&   508&   710&  1130\\
Slope, low $k/k_0$
&$ -0.77\pm  0.04$ 
&$ -0.73\pm  0.03$ 
&$ -0.75\pm  0.03$ 
&$ -0.76\pm  0.02$ 
&$ -0.58\pm  0.02$\\
Slope, high $k/k_0$
&$ -1.17\pm  0.01$ 
&$ -1.28\pm  0.02$ 
&$ -1.32\pm  0.03$ 
&$ -1.36\pm  0.03$ 
&$ -1.45\pm  0.03$\\
Break scale (pc)\tablenotemark{b} &$148_{17}^{30}$ &$118_{10}^{20}$ &$114_9^{18}$ &$113_9^{17}$ &$126_{12}^{35}$\\
&&&&&\\
\enddata
\tablenotetext{a}{Peak intensity threshold in units of $10^{-13}$ erg s$^{-1}$ cm$^{-2}$ arcsec$^{-2}$.}
\tablenotetext{b}{Break scales are from the intersection points of the power law fits to the PS on scales larger and smaller than $\sim100$ pc. The slopes of the PS are from fits at $k/k_0$ smaller than and larger than 120 pc, which is about the average of the tabulated break scales.}
\end{deluxetable*}

\begin{deluxetable*}{cccccc}
\tabletypesize{\scriptsize}
\tablecaption{Power Spectrum Results for H$\alpha$ at Various Radii\label{table:radii}}
\tablewidth{0pt}
\startdata
&&&&&\\
&&&&&\\
&\multicolumn{5}{c}{$16^{\prime\prime}$ to $35^{\prime\prime}$ (0.58-1.27 kpc)}\\
&&&&&\\
\hline
&&&&&\\
Threshold\tablenotemark{a}&  0.216&  0.324&  0.432&  0.648&none\\
Number of scans&   21&    51&    73&   102&   128\\
low $k/k_0$
&$ -0.97\pm  0.04$ 
&$ -0.96\pm  0.03$ 
&$ -0.95\pm  0.04$ 
&$ -0.91\pm  0.04$ 
&$ -0.77\pm  0.03$\\
high $k/k_0$
&$ -2.24\pm  0.34$ 
&$ -2.35\pm  0.18$ 
&$ -2.53\pm  0.16$ 
&$ -2.60\pm  0.15$ 
&$ -2.71\pm  0.11$\\
Break scale\tablenotemark{b}
&$29.3_{2.6}^{7.0}$ &$28.3_{2.3}^{6.5}$ &$26.5_{2.0}^{6.0}$ &$25.7_{1.8}^{5.8}$ &$25.7_{2.1}^{8.8}$\\
&&&&&\\
\hline
&&&&&\\
&\multicolumn{5}{c}{$35^{\prime\prime}$ to $80^{\prime\prime}$ (1.27-2.91 kpc)}\\
&&&&&\\
\hline
&&&&&\\
Threshold\tablenotemark{a}& 0.108&  0.162&  0.216&  0.324&none\\
Number of scans&    76&   143&   187&   239&   347\\
low $k/k_0$
&$ -0.87\pm  0.04$ 
&$ -0.82\pm  0.04$ 
&$ -0.81\pm  0.04$ 
&$ -0.79\pm  0.04$ 
&$ -0.57\pm  0.03$\\
high $k/k_0$
&$ -1.23\pm  0.02$ 
&$ -1.39\pm  0.02$ 
&$ -1.48\pm  0.03$ 
&$ -1.58\pm  0.04$ 
&$ -1.48\pm  0.06$\\
Break scale\tablenotemark{b}
&$110_7^{11}$
&$87_4^8$
&$81_5^9$
&$77_6^{13}$
&$90_9^{29}$\\
&&&&&\\
\hline
&&&&&\\
&\multicolumn{5}{c}{$80^{\prime\prime}$ to $180^{\prime\prime}$ (2.91-6.54 kpc)}\\
&&&&&\\
\hline
&&&&&\\
Threshold\tablenotemark{a}& 0.108&  0.162&  0.216&  0.324&none\\
Number of scans&   49&   143&   225&   372&   680\\
low $k/k_0$
&$ -0.77\pm  0.04$ 
&$ -0.71\pm  0.03$ 
&$ -0.75\pm  0.03$ 
&$ -0.74\pm  0.03$ 
&$ -0.55\pm  0.03$\\
high $k/k_0$
&$ -1.10\pm  0.02$ 
&$ -1.28\pm  0.01$ 
&$ -1.35\pm  0.02$ 
&$ -1.44\pm  0.02$ 
&$ -1.50\pm  0.04$\\
Break scale\tablenotemark{b} 
&$218_{26}^{42}$
&$192_{25}^{54}$
&$171_{17}^{34}$
&$171_{15}^{34}$
&$173_{13}^{41}$\\
&&&&&\\
\enddata
\tablenotetext{a}{Peak intensity threshold in units of $10^{-13}$ erg s$^{-1}$ cm$^{-2}$ arcsec$^{-2}$.}
\tablenotetext{b}{Break scales are from the intersection points of the power law fits to the PS on scales larger and smaller than characteristic values of $\sim20$ pc, $\sim80$ pc and $\sim150$ pc, in order of increasing galactocentric radii. The slopes of the PS are from fits at $k/k_0$ smaller than and larger than the averages of the fitted break scales, which are taken to be 25 pc, 90 pc and 180 pc, respectively. }
\end{deluxetable*}

\begin{deluxetable*}{cccccc}
\tabletypesize{\scriptsize}
\tablecaption{Power Spectrum Results for Pa$\alpha$ at Various Peak Intensity Thresholds\label{table:slopespa}}
\tablewidth{0pt}
\startdata
&&&&&\\
&&&&&\\
Threshold\tablenotemark{a}&0.38&0.51&0.64&1.0&2.0\\
Number of scans &    43&    69&    89&   111&   134\\
Slope, low $k/k_0$
&$ -0.61\pm  0.07$ 
&$ -0.64\pm  0.07$ 
&$ -0.65\pm  0.07$ 
&$ -0.70\pm  0.08$ 
&$ -0.64\pm  0.08$\\
Slope, high $k/k_0$
&$ -0.83\pm  0.02$ 
&$ -0.92\pm  0.02$ 
&$ -0.98\pm  0.02$ 
&$ -1.02\pm  0.03$ 
&$ -1.00\pm  0.03$\\
Break scale\tablenotemark{b}  &$168_{42}^{81}$
&$195_{40}^{76}$
&$183_{34}^{66}$
&$121_{25}^{46}$
&$113_{22}^{45}$\\
&&&&&\\
\enddata
\tablenotetext{a}{Peak intensity threshold in units of $10^{-13}$ erg s$^{-1}$ cm$^{-2}$ arcsec$^{-2}$.}
\tablenotetext{b}{Break scales are from the intersection points of the power law fits to the PS on scales larger and smaller than a characteristic value of $\sim200$ pc. The slopes of the PS are from fits at $k/k_0$ smaller than and larger than the average of the fitted break scale, which is $\sim170$ pc. }

\end{deluxetable*}

\begin{deluxetable*}{ccccc}
\tabletypesize{\scriptsize}
\tablecaption{Power Spectrum Results for Broadband Mid-Infrared Images\label{table:slopesBB}}
\tablewidth{0pt}
\tablehead{
\colhead{}&\multicolumn{4}{c}{JWST Filter}}
\startdata
&&&&\\
&F560W&F770W&F1000W&F1130\\
Threshold\tablenotemark{a}&20&    95&    35&   110\\
Number of scans& 26&    26&    33&    33\\
Slope, low $k/k_0$&$ -0.95\pm  0.09$ & $ -0.74\pm  0.08$ & $ -0.92\pm  0.06$ & $ -0.80\pm  0.06$\\
Slope, high $k/k_0$&$ -1.40\pm  0.21$ & $ -2.13\pm  0.42$ & $ -1.72\pm  0.24$ & $ -1.91\pm  0.55$\\
Break scale\tablenotemark{b} 
&$25.8_{4.6}^{8.7}$
&$31.8_{3.8}^{14}$
&$31.0_{3.3}^{7.3}$
&$36.3_{3.8}^{11}$\\
&&&&\\
\hline
\colhead{}&\multicolumn{4}{c}{JWST Filter}\\
\hline
&&&&\\
&F1280W&F1500W&F1800W&F2100W\\
Threshold\tablenotemark{a}&    90&    65&   100&   130\\
Number of scans&     33&    32&    30&    35\\
Slope, low $k/k_0$&$ -0.78\pm  0.07$ & $ -0.88\pm  0.07$ & $ -0.91\pm  0.06$ & $ -0.93\pm  0.07$\\
Slope, high $k/k_0$&$ -1.87\pm  1.39$ & $ -1.58\pm  5.20$ & ... & ...\\
Break scale\tablenotemark{b}  
&$34.7_{4.0}^{12}$
&$35.4_{4.7}^{10.4}$
&...
&...\\
&&&&\\
\enddata
\tablenotetext{a}{Peak intensity threshold in units of MJy sr$^{-1}$.}
\tablenotetext{b}{The slopes of the PS are from fits at $k/k_0$ smaller than and larger than a fiducial break scale of 100 pc.  The tabulated break scales are 
from the intersection points of the resulting power law fits. There are no evident breaks in the broadband PS, so these scales are not considered to be measures of disk thickness.}
\end{deluxetable*}

\vspace{0.5cm}

{\it Acknowledgements:} 
Helpful comments by Dr. Alex Lazarian and the referee, Dr. Frederic Bournaud, are appreciated. This work is based on observations made with the NASA/ESA/CSA James Webb Space Telescope and retrieved from the Mikulski Archive for Space Telescopes at the Space Telescope Science Institute, which is operated by the Association of Universities for Research in Astronomy, Inc., under NASA contract NAS 5-03127. These observations are associated with GO programs \# 1783 and  \#3435. The work also made use of archival data from the NASA/ESA Hubble Space Telescope obtained from the Space Telescope Science Institute, which is operated by the Association of Universities for Research in Astronomy, Inc., under NASA contract NAS 5–26555. Data presented in this article were obtained from the Mikulski Archive for Space Telescopes (MAST) at the Space Telescope Science Institute. The specific observations analyzed can be accessed via \dataset[DOI:10.17909/jp5q-s259]{doi.org/10.17909/jp5q-s259} (Dataset Title: M51/Elmegreen/Calzetti).
Support for programs GO \# 1783 and  GO \#3435 was provided by NASA through grants from the Space Telescope Science Institute. VJ and BG acknowledge support from grant JWST-GO-01783; K.S., D.D., M.L.B., D.C., A.K.L., and J.D.S. acknowledge support from grant JWST-GO-03435. A.A. acknowledges support from the Swedish National Space Agency (SNSA) through grant 2021-00108. R.S.K. acknowledges financial support from the European Research Council via the ERC Synergy Grant ``ECOGAL'' (project ID 855130),  from the German Excellence Strategy via the Heidelberg Cluster of Excellence (EXC 2181 - 390900948) ``STRUCTURES'', and from the German Ministry for Economic Affairs and Climate Action in project ``MAINN'' (funding ID 50OO2206). R.S.K. also thanks the 2024/25 Class of Radcliffe Fellows for their company and for highly interesting and stimulating discussions. K.G. is supported by the Australian Research Council through the Discovery Early Career Researcher Award (DECRA) Fellowship (project number DE220100766) funded by the Australian Government. A.D.C. acknowledges the support from the Royal Society University Research Fellowship URF/R1/191609.

{\it Facility:} HST, JWST (NIRCam, MIRI). Software: JWST Calibration Pipeline \citep{bushouse22, greenfield16}; SAOImage DS9 \citep{joye03}; FORTRAN, and VossPlot \citep{voss95}.


\end{document}